\documentstyle[11pt]{article}

\newcommand{\nl}{\nonumber \\}
\newcommand{\be}{\begin{equation}}
\newcommand{\ee}{\end{equation}}
\newcommand{\ba}{\begin{eqnarray}}
\newcommand{\ea}{\end{eqnarray}}
\newcommand{\Lgg}[1]{\ln^{#1}\frac{\Lambda^2}{M_0^2}}
\newcommand{\lgg}[1]{\ln^{#1}\frac{\Lambda^2}{\mu^2}}
\newcommand{\acc}[1]{\left[
1+ O\left(\frac{1}{\ln^{#1}\frac{\Lambda^2}{\mu^2}}\right)\right]}

\begin{document}
\addtolength{\textwidth}{1.8cm}
\begin{center}
\large \bf A.A.Andrianov, V.A.Andrianov, V.L.Yudichev
\end{center}
\ \vspace{1cm}\\
\begin{center}
{\Large\bf  Dynamical P-parity Breaking in Effective Quark
Model. }\\
{\it Department of Theoretical Physics, Sankt-Petersburg
 University.  }
\end{center}
\ \vspace{1cm}\\
{\renewcommand{\baselinestretch}{0.7}
\small
 Fermion models with selfinteraction including
derivatives of fields are investigated in the strong coupling regime. The
existence of three phases is
 established in the two channel model near tricritical point.
The special phase  of dynamical P-parity breaking is found. }

\section{ Introduction.}
\hspace{5mm}Quark models with local four-fermion interaction are widely
applied for the description of  low-energy  phenomena in the particle
and nuclear physics \cite{Eguchi}.
In the extended version these models can be supplemented with
 quasilocal interaction, including derivatives of fermion
fields~\cite{AAA}.
 For such models in the strong coupling regime the dynamical chiral
symmetry breaking occurs and it results in  the formation
of dynamical quark mass as well as in the production of
several meson states with the same quantum
numbers~\cite{Zvenigorod} (so called "radial excitations"~
\cite{PartData}).

It seemed to be the general feature for
the quark models of Nambu-Jona-Lasinio type that, when being a
  symmetry of the quark lagrangian, the P-parity remains a good
quantum number after the DCSB.  However, it happens that for
particular combination of four-fermion coupling constants in
  quasilocal vertices the P-parity can be broken dynamically
together with the chiral symmetry.

In our talk we focus on
composite boson states (mesons) created in scalar and
pseudoscalar channels due to  DCSB.

The two-channel model is
examined in details near tricritical point and three major
phases are described: a symmetrical one and two phases with
DCSB, different in correlation lengths in scalar channels.
On a particular plane in the coupling constant space the complex
solution for the dynamical mass function is found, which yields
the (logarithmically suppressed) P-parity breaking in  meson
sector.  It means that in such a phase there exist heavy scalar
states which can decay into  two or three pions.

\section{ Formulation of Quasilocal NJL-like Model.}

\hspace{5mm}We consider  a class of effective fermion models
with quasilocal interaction~\cite{Zvenigorod}.
The lagrangian for a minimal model which allows
the strong coupling regime in $l$-channels is following:

\be
{\cal L}(x)=i\bar\psi\partial_{\mu}\gamma^{\mu}\psi
+\frac{8\pi^2}{N_c\Lambda^2}\sum_{i,j=1}^l
a_{ij}\bar\psi_L\varphi_i\biggl(-
\frac{\partial^2}{\Lambda^2}\biggr)\psi_R\bar\psi_R\varphi_j\biggl(-
\frac{\partial^2}{\Lambda^2}\biggr)\psi_L.              \label{3}
\ee
Here $\psi_{L(R)}\equiv 1/2(1\pm\gamma_5)\psi$ are left and right
spinors; $a_{ij}$ is a non-degenerate real matrix
 of coupling constants.
The singlet flavor fermionic fields in the
model are transformed under $ SU_c(N_c)\times U_F(1) $
group, where $ N_c\gg 1 $ is the number of colors, and $ U_F(1)
$ reflects the flavor symmetry.
The sum over color indices is assumed.
The formfactors $\varphi_i(\tau),(\tau \rightarrow
-\partial^2/\Lambda^2)$ determine quasilocal
interactions. We regularize the vertices by imposing the
momentum cut-off $ \Lambda $ :
$\bar\psi\psi\longrightarrow\bar\psi\theta(\Lambda^2+\partial^2)\psi$,
and choose the formfactors $ \varphi_i $ to implement
\be
\int\limits_{0}^{1} d\tau \varphi_i(\tau)\varphi_j(\tau)=\delta_{ij} .
\label{6}
\ee

In order to develop the  large-$ N_c $ approximation,
 it is convenient to express~(\ref{3}) in
terms of auxiliary fields $ \chi_i=\sigma_i+i\pi_i $:
\be
{\cal L}_{\chi}(x)
=i\sigma_j\bar\psi\varphi_j\psi-\pi_j\bar\psi\varphi_j
\gamma_5\psi+\frac{N_c\Lambda^2}{8\pi^2}
\chi^{\star}_ia_{ij}^{-1}\chi_j . \nonumber
\ee
Thereby, we come to a model with
{\it l}\  scalar and pseudoscalar channels. When integrating
out the fermionic fields, we obtain the effective action of
$\chi,\chi^{\star}$:
 \be
 \exp(-S_{eff}(\chi))=<\exp(-\int d^4x {\cal
L}_{\chi}(x))>_{\psi,\bar\psi} \label{8} \ee ( in the Euclidian
space-time ).

The effective potential, $V_{eff}=S_{eff}\cdot
(\mbox{vol})^{-1}$, proves to be a functional depending on
the dynamical mass function $M(\tau)\equiv
\chi_j\varphi_j(\tau)$ and proportional to $ N_c $.  It allows
us to use the saddle point approximation for $N_c\gg 1$.  The
extrema of the effective potential can be found from mass-gap
equation:

\be
\frac{\delta V_{eff}}{\delta \chi_j^{\star}}=0. \label{9}
\ee
The trivial solution $ \chi_j=0 $ satisfies (\ref{9}), and
relates to the symmetrical phase. Non-trivial solutions bring
a non-zero dynamical fermion mass  $<M>_{\chi}\not=0$.
 We study the model near  the polycritical
point, where $M\ll \Lambda$:

\be
      a_{ij}\sim\delta_{ij}+\frac{\Delta_{ij}}{\Lambda^2}, \qquad
      |\Delta_{ij}|\ll \Lambda^2                            \label{11}
      \ee
when the strong coupling regime occurs in each of the
$l$-channels.

\section{ Two-Channel Model with DCSB.}

\hspace{5mm}A generalized NJL-like model with tricritical point is produced by
two-channel interaction (\ref{3}), when we set $i,j=1,2$ in
(\ref{3})-(\ref{11}).
We retain only the lowest derivatives in the potential, with
$\varphi_1=1$,  $\quad \varphi_2=\sqrt{3}(1-2\tau)$. The dynamical
mass function is thereby
$M(\tau)=\bar\chi_1+\bar\chi_2\sqrt{3}(1-2\tau)$.  As $\bar \chi_j$
are complex functions, $M(\tau)$ is complex too. However, with the global
chiral rotation $M(\tau)\rightarrow M(\tau)e^{i\omega}, \quad \omega=const$
it is always possible to implement $\mbox{Im}<M_0>_{\chi}=0$ and we
can choose the following parametrization:
\be
\bar\chi_1=\chi_1+i\rho, \quad
\bar\chi_2=\chi_2-i\frac{\rho}{\sqrt{3}},\quad \chi_i\equiv\mbox{ Re}
\bar\chi_i .
\label{13}
\ee
The  equations~(\ref{9}) for the two-channel model read
\ba
\Delta_{11}\chi_1+\Delta_{12}\chi_2&=&M_0^3\Lgg{} -6\sqrt{3}\chi_1^2\chi_2-
18\chi_1\chi_2^2-8\sqrt{3}\chi_2^3\nl
d_1\chi_1-d_2\chi_2&=&2\sqrt{3}\chi_1(\chi_1^2+3\chi^2_2)+2\rho^2(
\frac{4}{\sqrt{3}}\chi_1-2\chi_2)\nl
\rho(\sqrt{3}\Delta_{11}-\Delta_{12})&=&2\rho\sqrt{3}(\chi_1^2+\chi_2^2+
\frac43\rho^2), \label{14}
\ea
where
\be
d_1=\sqrt{3}\Delta_{11}-\Delta_{12},\quad\quad
d_2=-\sqrt{3}\Delta_{21}+\Delta_{22} \label{15}.
\ee

We analyze the equations (\ref{14}) near  polycritical point,
$|\Delta_{ij}|\sim {\mu}^2\ll \Lambda^2$, in the large-log
approximation ($\ln\frac{\Lambda^2}{{\mu}^2}\gg \ln\ln
\frac{\Lambda^2}{{\mu}^2}$). It gives rise to a set of classes of
solutions.

For $\rho=0$ all the solutions are divided to the following classes:

  a) Gross-Neveu-like solutions $\chi_j^{GN}$:

\be
\chi^2_1=\frac{d_2^2\det
\Delta}{(\sqrt{3}d_1+d_2)^3\lgg{}}\acc{},\qquad
\chi_2\approx\frac{d_1}{d_2}\chi_1.
  \label{17}
\ee
These solutions deliver  minima to the potential when
$\sqrt{3}d_1+d_2 < 0$, with one  eigenvalue of the matrix $\Delta$
being in the overcritical regime, the other in the undercritical.

  b) Abnormal solutions are:

\be
\chi^2_1=\frac{\sqrt{3}d_1+d_2}{12}\acc{1/3}, \qquad
\chi_2\approx -\frac{\chi_1}{\sqrt{3}}.  \label{19}
\ee
They give  minima to the potential when
$\sqrt{3}d_1+d_2 >0$, $\sqrt{3}d_1-2d_2\not =0$ ( either both
eigenvalues  of $\Delta$ are positive, or one is positive and
the other negative ).

c) On the hyperplanes $\sqrt{3}d_1+d_2=0$ and
$\sqrt{3}d_1-2d_2=0$ there appear special solutions with a
different asymptotics~\cite{Zvenigorod}.
 In general, in the models with more than one channel
the complex solutions are allowed, and the imaginary parts of
all the variables $\chi_j$ cannot be removed simultaneously by
a global chiral rotation.

d) Complex solutions ($\rho\not=0$) appear for those domains
in vicinity of the hyperplane
$\sqrt{3}d_1-2d_2=0$, their asymptotic expressions are:

\be
\chi_1^2=\frac{d_1+4\Delta_{12}}{16\sqrt{3}(\lgg{}-3)} \qquad
\chi_2\approx -\sqrt{3}\chi_1,  \label{31}
\ee
and the dynamical mass is $M_c^2=4\chi_1^2$. The axial part of
the mass function looks as follows:

\be
\rho^2=\frac{d_1\sqrt{3}}{8}-\frac34( \chi^2_1+\chi^2_2 )=
\frac{d_1\sqrt{3}}{8}
\acc{}.                             \label{31'}
\ee

In each of the phase space's domains  mentioned above one finds
four common boson states --- two scalar and two pseudoscalar
--- for real $\chi_j$, and, in general, three states with mixed
P-parity and one pseudoscalar with zero mass, the latest is in
accordance to the Goldstone theorem. We discuss the spectrum of
revealed states  in the next section.

\section{Second Variation and Mass Spectrum of Composite
States.}

\hspace{5mm}The matrix of  second
variations of the effective potential
determines the spectrum of bosonic states. We divide it on two
parts: one independent on momentum, $\hat B$, and the kinetic part,
which  is proportional to momentum squared, $\hat A p^2$.
\be
\delta^{(2)}S=(\delta\chi^{\star}, (\hat Ap^2+\hat B)\delta\chi)
\ee
\be
\chi_j=<\!\chi_j\!>+\delta\chi_j =<\!\chi_j\!>+
\sigma_j+i\pi_j.
\ee
The constant matrix $\hat B$ has zero-mode
$ \chi^0_j=<\!\pi_j\!>\!- i\cdot
<\!\sigma_j\!> $, regarding to the existence of Goldstone
bosons.

To find the spectrum of collective excitations one should solve
the equation
\be
\label{41} \det( \hat Ap^2+\hat B )=0.
\ee
Taking into account
the conditions necessary for a minimum of the potential,
 we find the solutions at $-m^2=p^2\le 0$,
giving physical values  of particle masses.

In the case of $\rho=0$:

  a) NJL-like mass spectrum:

\ba
&& m_{\pi}^2=0\quad m_{\pi'}^2\approx m_{\sigma'}^2\approx
-\frac{\sqrt{3}d_1+d_2}{3}\nl
&&m^2_{\sigma}\approx 4M_0^2,     \label{42}
\ea
 In this
domain  the radial excitation states are heavier than the lightest
scalar meson by a factor of logarithm.

  b) For the abnormal solutions we have

\ba
m_{\pi}^2=0,&& m_{\pi'}^2\approx \frac19\left( \frac43 \right)^{1/3}
\frac{( \sqrt{3}d_1-2d_2)^{4/3}}{( \sqrt{3}d_1+d_2 )^{1/3}}
\frac{1}{\lgg{1/3}},
\label{44} \\
m_{\sigma}^2\approx 6M_{an}^2&&m_{\sigma'}\approx
\frac23( \sqrt{3}d_1+d_2 ).\nonumber
\ea

When comparing (\ref{42}) and (\ref{44}) we find the scalar
channel correlation length to be different for each phase, that
corresponds to the tricritical point conditions.

c) For the
special real solutions the relations between scalar and
pseudoscalar meson masses are different
from (\ref{42}),(\ref{44}) (see~\cite{Zvenigorod}).

 \section{Mass Spectrum in the P-parity Breaking Phase.}
\hspace*{5mm}One can see from (\ref{31}),(\ref{31'}) that in
the large-log approximation the axial dynamical mass
(the imaginary part of $M(\tau)$) dominates. It leads to
appearance of a massless boson in the scalar channel
in accordance to the Goldstone theorem. Conventionally, the
massless boson is interpreted as  $\pi$-meson. For this purpose
we make a global chiral rotation of fermionic fields $
\psi\rightarrow \exp( i\gamma_5\pi/4 )\psi $ accompanied by
corresponding rotation of the bosonic variables $\bar \chi_j
\rightarrow i\bar \chi_j $:

\be
\bar\chi_1=i\chi_1-\rho,\quad \bar\chi_2=i\chi_2+\frac{\rho}{\sqrt{3}}
\label{57}
\ee
The classification of states given by P-parity quantum
number is relevant only in the large-log approximation,
when
\be
\frac{B^{\pi\sigma}}{B^{\sigma\sigma}}\approx
\frac{B^{\pi\sigma}}{B^{\pi\pi}}=O\left( \frac{1}{\lgg{}}
\right) \label{58}
\ee
next-to-leading logarithmic
effects are of no importance and one can neglect mixing of the
states with different P-parity.
Then the
spectrum of mesons is:

\begin{eqnarray}
  m_{\pi}^2=0,&& m_{\pi'}^2\approx \frac{d_1+4\Delta_{12}}{\sqrt{3}\lgg{}}
\approx 16\chi_1^2=4M_c^2
\label{59}      \nl
m_{\sigma'}^2\approx \sqrt{3}d_1,&& m_{\sigma}^2\approx
\frac{4( d_1+\Delta_{12} )}{9\sqrt{3}\lgg{}}
\end{eqnarray}
The ratio of $m_{\pi'}$ and $m_{\sigma}$
does not depend on logarithm, so both the masses are comparable.
On the other hand, in the
models with finite momentum cut-off,
the effects of order of
$ \Lambda^2 $ become sensible,
 the dynamical P-parity
breaking is induced, since $ B_{\pi\sigma}\not=0$.

This phenomenon  of dynamical P-parity breaking can be used in
extensions of the Standard Model~\cite{Miransky} where several
Higgs bosons are composite ones.
Thus we conclude that the models with polycritical points are
drastically different from the local NJL models in the variety
of the physical phenomena in the DCSB.

We are indebted to the Organizing Committee of the X Workshop on
HEP and QFT and especially to Prof. V.Sarvin for hospitality, many informal
and fruitful discussions and for opportunity to present our report.
This report is supported by the RFFI Grant
No. 95-02-05346-a and Grant INTAS-93-283.

\end{document}